\documentclass[conference]{IEEEtran}

\usepackage{graphicx}
\newtheorem{theorem}{\bf Theorem}
\newtheorem{corollary}{\bf Corollary}

\newtheorem{definition}{\it Definition}
\newtheorem{example}{\bf Example}

\begin{document}

\sloppy

\title{Distributed Capacity of A Multiple Access Channel}

\author{
  \IEEEauthorblockN{Yanru Tang, Faeze Heydaryan, and Jie Luo}
  \IEEEauthorblockA{Electrical \& Computer Engineering Department \\
  Colorado State University, Fort Collins, CO 80523 \\
  Email: \{yrtang, faezeh66, rockey\}@colostate.edu}
}



\maketitle

\begin{abstract}
In distributed communication, each transmitter prepares an ensemble of channel codes. To encode a message, a transmitter chooses a channel code individually without sharing the coding choice with other transmitters or with the receiver. Upon receiving the block of channel output symbols, the receiver either decodes the messages of interest if a pre-determined reliability requirement can be met, or reports collision otherwise. Revised from the distributed channel coding theorems in the literature, distributed capacity of a discrete-time memoryless multiple access channel is defined and derived under the assumption that codeword length can be taken to infinity. An improved achievable error performance bound is presented for the case when codeword length is finite.\footnote{This work was supported by the National Science Foundation under Grants CCF-1420608 and CNS-1618960. Any opinions, findings, and conclusions or recommendations expressed in this paper are those of the authors and do not necessarily reflect the views of the National Science Foundation.}
\end{abstract}

\section{Introduction}
Classical channel coding assumes that users in a communication party should jointly optimize their channel codes, and transmit encoded messages to the receiver over a long time duration. Overhead of achieving the required user coordination is often ignored based on the fundamental assumption that coordinated message transmission should dominate the communication process. However, this assumption is increasingly challenged by the dynamic packet-based communication activities in data networks. In a wireless network, not only messages can be short and bursty, coordinating a large number of users can also be expensive or infeasible in terms of overhead. A significant proportion of messages in existing wireless networks such as Wi-Fi systems are transmitted using distributed protocols where users make their communication decisions individually. Featured by opportunistic channel access and occasional packet collision, the distributed communication model does not fall into the classical channel coding framework. Its fundamental limits therefore cannot be understood without extending the classical channel coding tools.

Distributed channel coding theory, proposed in \cite{ref Luo12}\cite{ref Wang12}\cite{ref Luo15}, assumes that each transmitter should be equipped with an ensemble of channel codes as opposed to one code. Code ensembles are shared off-line with the receiver, e.g., by specifying codebook generation algorithms in the physical layer protocol. Different codes can correspond to different communication settings such as different rate and power combinations. During online communication, possibly depending on a link layer decision, each transmitter individually chooses a code to encode a messages. Without knowing the coding choices of the users, a receiver either decodes the messages of interest if a pre-determined decoding reliability requirement can be met, or reports collision otherwise. An achievable region is defined in \cite{ref Luo12}\cite{ref Luo15} as the set of code index vectors that support asymptotic reliable message recovery, and was shown to coincide with the Shannon information rate region in a sense explained in \cite{ref Luo12}\cite{ref Luo15}. Error performance bounds in the case of finite codeword length were obtained in \cite{ref Wang12}\cite{ref Luo15}. While fundamental understandings about distributed communication are much needed for packet-based wireless networks, coding theory developed in \cite{ref Luo12}\cite{ref Wang12}\cite{ref Luo15} has not been attracting much attention in the research community so far.

In this paper, we present two simple extensions to the distributed channel coding theorems obtained in \cite{ref Luo12}\cite{ref Wang12}\cite{ref Luo15}. First, in \cite{ref Luo12}\cite{ref Luo15}, achievable regions were defined not only as a function of the communication channel, but also as a function of the code ensembles selected by the users. We revise the definition to one that only depends on the communication channel. Such a revision enabled the definition of the distributed channel capacity, which is supported by the existing achievability proof and a new but quite straightforward converse proof. Second, error probability in a communication system is often dominated by a small number of error event types. In a distributed communication system, different error event types may or may not correspond to different code index vectors of the users. In \cite[Theorem 3]{ref Luo15}, the obtained achievable error performance bound contains a term that equals the probability of the worst case error event type multiplies the number of code index vectors outside the operation region. If the latter parameter takes a large value, the corresponding error performance bound can be very loose. We revise the derivation to obtain a performance bound that essentially replaces the particular term with a summation of error probabilities each corresponding to one code index vector. The new error performance bound is tighter than the one obtained in \cite{ref Luo15} because the new bound is unlikely to scale in the number of code index vectors.

\section{Multiple Access with Single User Decoding}
\label{SectionII}
Consider a multiple access system with $K$ transmitters (users) and one receiver. Time is slotted with each time slot equaling the length of $N$ channel symbols, and this is also the length of a codeword. Throughout the paper, we only consider channel coding within one time slot. We use bold font variable to represent a vector whose entries are the corresponding variables of all users. The discrete-time memoryless channel is modeled by a conditional distribution $P_{Y|\mbox{\scriptsize \boldmath  $X$}}$, where $\mbox{\boldmath  $X$}=[X_1, \dots, X_K]\in \mbox{\boldmath  $\mathcal{X}$}$ is the channel input symbol vector with $\mbox{\boldmath  $\mathcal{X}$}$ being the vector of finite input alphabets, and $Y\in \mathcal{Y}$ is the channel output symbol with $\mathcal{Y}$ being the finite output alphabets. We assume that channel input alphabet $\mathcal{X}_k$ should be known at user $k$, for $k=1, \dots, K$, and the conditional distribution $P_{Y|\mbox{\scriptsize \boldmath  $X$}}$ should be known at the receiver.

Each transmitter, say user $k$, is equipped with an ensemble of $M$ channel codes, denoted by $\mathcal{G}_k^{(N)}=\{g_{k1}, \dots, g_{kM}\}$. Let $\mbox{\boldmath  $\mathcal{G}$}^{(N)}$ denote the vector of code ensembles of all users. Let $\mbox{\boldmath  $g$}=[g_1, \dots, g_K]$ be a code index vector. We say $\mbox{\boldmath  $g$}\in \mbox{\boldmath  $\mathcal{G}$}^{(N)}$ if $g_k\in \mathcal{G}_k^{(N)}$ for all $1\le k\le K$. For each user $k$, each index $g_k\in \mathcal{ G}_k^{(N)}$ represents a random block code described as follows. Let $\mathcal{L}_{g_k}=\left\{\mathcal{C}_{g_k\theta_k}: \theta_k \in \Theta_k^{(N)}\right\}$ be a library of codebooks, indexed by a set $\Theta_k^{(N)}$. Each codebook contains $\lfloor e^{Nr_{g_k}} \rfloor$ codewords of length $N$, where $r_{g_k}$ is a pre-determined parameter termed the ``communication rate'' (in nats/symbol) of code $g_k$. Let $[\mathcal{C}_{g_k\theta_k}(w_k)]_j$ denote the $j$th symbol of the codeword corresponding to message $w_k$ in codebook $\mathcal{C}_{g_k\theta_k}$. At the beginning of each time slot, a codebook index $\theta_k$ is generated randomly according to a distribution $\gamma_k^{(N)}$. The distribution $\gamma_k^{(N)}$ and the codebooks $\mathcal{C}_{g_k\theta_k}$, $\forall g_k \in \mathcal{G}_k^{(N)}$, are chosen such that random variables $X_{g_kw_kj} : \theta_k \to [\mathcal{C}_{g_k\theta_k}(w_k)]_j$, $\forall j, w$ and $\forall g_k$, are i.i.d. according to a pre-determined input distribution $P_{g_kX_k}$. Assume that code library $\mathcal{L}_{g_k}$ and the value of $\theta_k$ are both known at the receiver. That is, the receiver knows the randomly generated codebook of $g_k$, and this is true for all codes and for all users. Note that this can be achieved by sharing the random codebook generation algorithms with the receiver. In the above description, a random block code $g_{k}$ is characterized by its communication rate $r_{g_k}$ and its input distribution $P_{g_kX_k}$. With an abuse of the notation, we regard $g_k=(r_{g_k}, P_{g_kX_k})$ as a variable representing a rate and distribution pair of user $k$, which is not a function of the codeword length $N$. Similarly, we regard $\mbox{\boldmath  $g$}=(\mbox{\boldmath  $r$}_{\mbox{\scriptsize \boldmath  $g$}}, \mbox{\boldmath $P$}_{\mbox{\scriptsize \boldmath $gX$}})$ as a vector variable representing the rate and distribution pairs of all users. We will use ``code space'' to refer to the space of $\mbox{\boldmath  $g$}$, which is also the space of rate vector and distribution vector pairs. We use $\mbox{\boldmath  $\mathcal{G}$}$, i.e., without superscription $(N)$, to represent a code ensemble in the code space where each $\mbox{\boldmath  $g$}\in \mbox{\boldmath  $\mathcal{G}$}$ represents a point in the code space.

At the beginning of each time slot, we assume that each user, say user $k$, arbitrarily chooses a code $g_k \in \mathcal{G}_k^{(N)}$, maps a message $w_k$ to a codeword $X_{g_k}^{(N)}(w_k)$, and then sends the codeword through the channel. Here ``arbitrary'' refers to the assumption that the coding choice is not controlled by, and even its statistical information may not be known to the physical layer transmitter. Assume $(\mbox{\boldmath $w$}, \mbox{\boldmath $g$})$ is the actual message vector and code index vector chosen by the transmitters. Let $\mbox{\boldmath $X$}_{\mbox{\scriptsize \boldmath $g$}}^{(N)}(\mbox{\boldmath $w$})$ be the vector of codewords. We assume that neither $\mbox{\boldmath $g$}$ nor $\mbox{\boldmath $w$}$ is known at the receiver.

We assume that the receiver is only interested in decoding the message of user $1$, but can choose to decode the messages of some other users if necessary. Because users choose their codes arbitrarily, reliable message decoding is not always possible. Upon receiving the channel output symbol sequence $Y^{(N)}=[Y_1, Y_2, \dots, Y_N]$, the receiver either outputs an estimated message and code index pair $(\hat{w}_1, \hat{g}_1)$ for user $1$, or reports collision for user $1$. We assume that the receiver should choose an ``operation region'' $\mbox{\boldmath $R$}_1$ in the code space. Without knowing the actual message vector and code index vector pair $(\mbox{\boldmath $w$}, \mbox{\boldmath $g$})$, the receiver intends to decode the message of user $1$ if $\mbox{\boldmath $g$}\in \mbox{\boldmath $R$}_1$, and intends to report collision for user $1$ if $\mbox{\boldmath $g$}\not \in \mbox{\boldmath $R$}_1$. Given the operation region $\mbox{\boldmath $R$}_1$ and conditioned on $\mbox{\boldmath $g$}$ being the actual code index vector, communication error probability as a function of $\mbox{\boldmath $g$}$ for codeword length $N$ is defined as follows.
\begin{eqnarray}
&& P_e^{(N)}(\mbox{\boldmath $g$})= \nonumber \\
&& \quad \left\{\begin{array}{ll}\max_{\mbox{\scriptsize \boldmath $w$}} Pr\{(\hat{w}_1, \hat{g}_1)\ne (w_1, g_1)|(\mbox{\boldmath $w$},\mbox{\boldmath $g$})\}, \forall \mbox{\boldmath $g$}\in \mbox{\boldmath $R$}_1 \\ \max_{\mbox{\scriptsize \boldmath $w$}} 1-Pr\left\{\left.\begin{array}{l}\mbox{``collision'' or}\\ (\hat{w}_1, \hat{g}_1)=(w_1, g_1) \end{array}\right|(\mbox{\boldmath $w$}, \mbox{\boldmath $g$}) \right\} \\ \hfill \forall \mbox{\boldmath $g$}\not\in \mbox{\boldmath $R$}_1\end{array} \right. \nonumber \\
\label{MultipleAccessSingleDecodingErrorProbability}
\end{eqnarray}
Note that in the above error probability definition, for $\mbox{\boldmath $g$}\not\in \mbox{\boldmath $R$}_1$, we regard both correct message decoding and collision report as acceptable channel outcomes. In other words, collision report is not strictly enforced for $\mbox{\boldmath $g$}\not\in \mbox{\boldmath $R$}_1$. A more general error probability definition will be discussed in Section \ref{SectionIV}.

\begin{definition}\label{MultipleAccessSingleDecodingAchievabilityDefinition}
	We say that an operation region $\mbox{\boldmath $R$}_1$ is asymptotically achievable for a multiple access channel $P_{Y|\mbox{\scriptsize \boldmath  $X$}}$ for user $1$, if for all finite $M$ and all code ensemble vectors $\mbox{\boldmath $\mathcal{G}$}$ with each entry of code ensemble having a cardinality of $M$, decoding algorithms can be designed for the sequence of random code ensembles $\mbox{\boldmath $\mathcal{G}$}^{(N)}$ to achieve $\lim_{N\to \infty} P_e^{(N)}(\mbox{\boldmath $g$})=0, \forall \mbox{\boldmath $g$}\in \mbox{\boldmath $\mathcal{G}$}.$	
\end{definition}

Compared with the achievable region definition given in \cite[Section III]{ref Luo15}, the achievable region defined in Definition \ref{MultipleAccessSingleDecodingAchievabilityDefinition} is only a function of the region and the multiple access channel. It does not depend on the particular code ensembles $\mbox{\boldmath  $\mathcal{G}$}$ chosen by the users. The following theorem is directly implied by the achievable region definition and the error probability definition given in (\ref{MultipleAccessSingleDecodingErrorProbability}).

\begin{theorem}\label{MultipleAccessSingleDecodingAchievability}
	For a discrete-time memoryless multiple access channel $P_{Y|\mbox{\scriptsize \boldmath  $X$}}$ with finite input and output alphabets, if an operation region $\mbox{\boldmath $R$}_1$ is asymptotically achievable for user $1$, then any subset $\tilde{\mbox{\boldmath $R$}}_1\subseteq \mbox{\boldmath $R$}_1$ is also asymptotically achievable for user $1$.
\end{theorem}

The following theorem characterizes the maximum achievable region of multiple access channel $P_{Y|\mbox{\scriptsize \boldmath  $X$}}$ for user $1$.

\begin{theorem}\label{MultipleAccessSingleDecodingCapacity}
 	For a discrete memoryless multiple access channel $P_{Y|\mbox{\scriptsize \boldmath  $X$}}$ with finite input and output alphabets, the following region $\mbox{\boldmath $C$}_{d1}$ in the code space is asymptotically achievable for user $1$.
 	\begin{equation}
 		\mbox{\boldmath $C$}_{d1}=\left\{\mbox{\boldmath $g$}\left| \begin{array}{l} \mbox{\boldmath $g$}=(\mbox{\boldmath $r$}_{\mbox{\scriptsize \boldmath $g$}}, \mbox{\boldmath $P$}_{\mbox{\scriptsize \boldmath $gX$}}), \forall S\subseteq \{1, \dots, K\}, 1\in S, \\ \exists \tilde{S}\subseteq S, 1\in \tilde{S}, \mbox{such that, }\\ \sum_{k\in \tilde{S}}r_{g_k} < I_{\mbox{\scriptsize \boldmath $g$}}(\mbox{\boldmath $X$}_{\tilde{S}};Y|\mbox{\boldmath $X$}_{\bar{S}}) \end{array} \right.\right\},
 		\label{MultipleAccessUser1Capacity}
 	\end{equation}
 	where $\bar{S}$ is the compliment set of $S$, $\mbox{\boldmath $X$}_{\bar{S}}$ is a vector of channel input symbols of users not in $S$, and $I_{\mbox{\scriptsize \boldmath $g$}}(\mbox{\boldmath $X$}_{\tilde{S}};Y|\mbox{\boldmath $X$}_{\bar{S}})$ denotes the mutual information between $\mbox{\boldmath $X$}_{\tilde{S}}$ and $Y$ given $\mbox{\boldmath $X$}_{\bar{S}}$ with respect to joint distribution $P_{\mbox{\scriptsize \boldmath $X$}Y}=P_{Y|\mbox{\scriptsize \boldmath $X$}}\prod_{k=1}^{K}P_{g_kX_k}$.
 	
 	The achievable region $\mbox{\boldmath $C$}_{d1}$ is maximum in the sense that for any region $\mbox{\boldmath $R$}_1$ that is asymptotically achievable for user $1$, we must have $\mbox{\boldmath $R$}_1\subseteq \mbox{\boldmath $C$}_{d1}^c$, where $\mbox{\boldmath $C$}_{d1}^c$ is the closure of $\mbox{\boldmath $C$}_{d1}$.
\end{theorem}

The proof of Theorem \ref{MultipleAccessSingleDecodingCapacity} is given in Appendix \ref{ProofofMultipleAccessSingleDecodingCapacity}.

Theorem \ref{MultipleAccessSingleDecodingCapacity} can be extended from decoding for a single user to decoding for a user subset.

 \begin{definition}
 	Let $S_0\subseteq \{1, \dots, K\}$ be a user subset. We say that an operation region $\mbox{\boldmath $R$}_{S_0}$ is asymptotically achievable for multiple access channel $P_{Y|\mbox{\scriptsize \boldmath  $X$}}$ for user subset $S_0$, if $\forall k\in S_0$, $\mbox{\boldmath $R$}_{S_0}$ is asymptotically achievable for user $k$.
 \end{definition}

 \begin{corollary}\label{MultipleAccessGroupDecodingCapacity}
 	For a discrete memoryless multiple access channel $P_{Y|\mbox{\scriptsize \boldmath  $X$}}$ with finite input and output alphabets, let $C_{dk}$ be the maximum achievable region for user $k$. The expression of $C_{dk}$ can be obtained from (\ref{MultipleAccessUser1Capacity}) by replacing user index $1$ with user index $k$. Let $S_0 \subseteq \{1, \dots, K\}$ be a user subset. The maximum achievable region for user subset $S_0$ is given by
 	\begin{eqnarray}
 		&& \mbox{\boldmath $C$}_{dS_0}= \bigcap_{k\in S_0}\mbox{\boldmath $C$}_{dk}= \nonumber \\
 		&& \left\{\mbox{\boldmath $g$}\left| \begin{array}{l} \mbox{\boldmath $g$}=(\mbox{\boldmath $r$}_{\mbox{\scriptsize \boldmath $g$}}, \mbox{\boldmath $P$}_{\mbox{\scriptsize \boldmath $gX$}}), \forall S\subseteq \{1, \dots, K\}, \\ S\cap S_0 \ne \phi, \exists \tilde{S}, S\cap S_0 \subseteq \tilde{S} \subseteq S, \\ \mbox{such that, } \sum_{k\in \tilde{S}}r_{g_k} < I_{\mbox{\scriptsize \boldmath $g$}}(\mbox{\boldmath $X$}_{\tilde{S}};Y|\mbox{\boldmath $X$}_{\bar{S}}) \end{array} \right.\right\},
 		\label{MultipleAccessUserS0Capacity}
 	\end{eqnarray}
 	where $\phi$ is the empty set.
 \end{corollary}

Corollary \ref{MultipleAccessGroupDecodingCapacity} can be obtained by following the proof of \cite[Theorem 4]{ref Luo12}.

Note that, according to \cite[Theorem 5]{ref Luo15}, Theorem \ref{MultipleAccessSingleDecodingCapacity} and Corollary \ref{MultipleAccessGroupDecodingCapacity} still hold even if we strictly enforce collision report for $\mbox{\boldmath $g$}\not\in \mbox{\boldmath $R$}_1$, by changing the error probability definition to
\begin{eqnarray}
&& P_e^{(N)}(\mbox{\boldmath $g$})= \nonumber \\
&& \quad \left\{\begin{array}{ll}\max_{\mbox{\scriptsize \boldmath $w$}} Pr\{(\hat{w}_1, \hat{g}_1)\ne (w_1, g_1)|(\mbox{\boldmath $w$},\mbox{\boldmath $g$})\}, \forall \mbox{\boldmath $g$}\in \mbox{\boldmath $R$}_1 \\ \max_{\mbox{\scriptsize \boldmath $w$}} 1-Pr\left\{\left.\mbox{``collision''} \right|(\mbox{\boldmath $w$}, \mbox{\boldmath $g$}) \right\} \hfill \forall \mbox{\boldmath $g$}\not\in \mbox{\boldmath $R$}_1\end{array} \right. \nonumber \\
\label{MultipleAccessSingleDecodingErrorProbability2}
\end{eqnarray}

With the support of Theorem \ref{MultipleAccessSingleDecodingCapacity} and Corollary \ref{MultipleAccessGroupDecodingCapacity}, we define $\mbox{\boldmath $C$}_{d1}$ as the ``distributed capacity'' for user $1$, and $\mbox{\boldmath $C$}_{dS_0}$ as the ``distributed capacity'' for user subset $S_0$, of multiple access channel $P_{Y|\mbox{\scriptsize \boldmath  $X$}}$. Interestingly, the distributed capacity can indeed be regarded as an extension to the classical Shannon capacity in the following sense.

Let $\mbox{\boldmath $C$}_d$ be the distributed capacity of the multiple access channel when the receiver is interested in decoding the messages of all users. According to Corollary \ref{MultipleAccessGroupDecodingCapacity}, $\mbox{\boldmath $C$}_d$ is given by
\begin{eqnarray}
&& \mbox{\boldmath $C$}_d=\Biggl\{ \mbox{\boldmath $g$} \Biggl| \mbox{\boldmath $g$}=(\mbox{\boldmath $r$}_{\mbox{\scriptsize \boldmath $g$}}, \mbox{\boldmath $P$}_{\mbox{\scriptsize \boldmath $gX$}}), \forall S\subseteq \{1, \dots, K\},   \nonumber \\
&& \quad \left. \sum_{k\in S}r_{g_k}<I_{\mbox{\scriptsize \boldmath $g$}}( \mbox{\boldmath $X$}_S;Y| \mbox{\boldmath $X$}_{\bar{S}})\right\}.
\label{DistributedCapacityMultipleAccess}
\end{eqnarray}
It is well known that Shannon capacity of the multiple access channel, denoted by $\mbox{\boldmath $C$}$, is given by
\begin{eqnarray}
&& \mbox{\boldmath $C$}=\mbox{convex hull}\left(\Biggl\{ \mbox{\boldmath $r$} \Biggl| \exists \mbox{\boldmath $P$}_{\mbox{\scriptsize \boldmath $X$}}, \forall S\subseteq \{1, \dots, K\},   \right. \nonumber \\
&& \quad \left.\left. \sum_{k\in S}r_k\le I( \mbox{\boldmath $X$}_S;Y| \mbox{\boldmath $X$}_{\bar{S}})\right\}\right),
\label{ShannonCapacityMultipleAccess}
\end{eqnarray}
where $I( \mbox{\boldmath $X$}_S;Y| \mbox{\boldmath $X$}_{\bar{S}})$ is calculated with respect to joint distribution $P_{\mbox{\scriptsize \boldmath $X$}Y}=P_{Y|\mbox{\scriptsize \boldmath $X$}}\prod_{k=1}^{K}P_{X_k}$. From (\ref{DistributedCapacityMultipleAccess}) and (\ref{ShannonCapacityMultipleAccess}), we can see that the two capacity terms satisfy
\begin{equation}
\mbox{\boldmath $C$}^c=\mbox{convex hull}\left(\{ \mbox{\boldmath $r$} | \exists \mbox{\boldmath $g$} \in \mbox{\boldmath $C$}_d^c, \mbox{\boldmath $r$}_{\mbox{\scriptsize \boldmath $g$}}=\mbox{\boldmath $r$}\}\right).
\end{equation}

Similar to classical channel coding theory, Theorem \ref{MultipleAccessSingleDecodingCapacity} and Corollary \ref{MultipleAccessGroupDecodingCapacity} hold even if input and output alphabets of the channel are continuous. One can also pose a constraint in the code space to limit the coding choices of the users, and to define the constrained distributed channel capacity accordingly.

\begin{example}
Consider a $K$-user multiple access system over a discrete-time memoryless channel with additive Gaussian noise. The channel is modeled by
\begin{equation}
Y = \sum_{k=1}^K X_k + V,
\end{equation}
where $V$ is the Gaussian noise with zero mean and variance $N_0$. Assume that each user $k$ can only choose random block codes with Gaussian input distribution of zero mean and variance $P_k$. With the input distributions being fixed, closures of the constrained distributed channel capacity and the Shannon capacity both equal the following rate region.
\begin{eqnarray}
&& \mbox{\boldmath $C$}_d^c=\mbox{\boldmath $C$}^c=\Biggl\{ \mbox{\boldmath $r$}_{\mbox{\scriptsize \boldmath $g$}} \Biggl| \forall S\subseteq \{1, \dots, K\},  \nonumber \\
&& \quad \left. \sum_{k\in S}r_{g_k} \le \frac{1}{2}\log\left(1+ \frac{\sum_{k\in S}P_k}{N_0} \right) \right\}.
\end{eqnarray}
However, the same capacity region has different meanings under different communication models. In coordinated communication, Shannon capacity region suggests that users should jointly choose a rate vector within the capacity region to guarantee reliable message delivery. In distributed communication, on the other hand, users choose their rates individually. If the rate vector happens to locate inside the capacity region, the receiver can detect it and decode the messages reliably. If the rate vector happens to locate outside the capacity region, the receiver can reliably detect it and report collision.
\end{example}

\section{Interfering User and Compound Channel}
\label{SectionIII}
In this section, we extend the coding theorems presented in Section \ref{SectionII} to the case when the system has an ``interfering user''. As explained in \cite{ref Luo15}, an interfering user can be a remote user whose codebook is unknown to the receiver, and hence its message is not decodable at the receiver. A ``virtual'' interfering user can also be used to model a compound channel whose realization affects the conditional channel distribution experienced by other users, but it is also ``virtual'' in the sense of having no message to be decoded at the receiver \cite{ref Luo15}.

Assume that, in addition to the $K$ regular users indexed by $\{1, \dots, K\}$, there is an interfering user indexed as user $0$. We assume that the interfering user is equipped with $M$ communication options, denoted by $\mathcal{G}_0=\{ g_{01}, \dots, g_{0M}\}$. For convenience, we still call $\mathcal{G}_0$ a code ensemble and call $g_0 \in \mathcal{G}_0$ a code index. With the existence of the interfering user, the multiple access channel is now modeled by a conditional distribution $P_{Y|\mbox{\scriptsize \boldmath  $X$}}(g_0)$, which is a function of the ``coding'' choice of the interfering user. Note that channel function $P_{Y|\mbox{\scriptsize \boldmath  $X$}}(g_0)$ can be defined for a domain of $g_0$ that is beyond the ensemble $\mathcal{G}_0$. At the beginning of each time slot, assume that the interfering user should arbitrarily choose a ``code'' $g_0$, and this determines the multiple access channel $P_{Y|\mbox{\scriptsize \boldmath  $X$}}(g_0)$ to be experienced by the regular users. The receiver knows the channel functions $P_{Y|\mbox{\scriptsize \boldmath  $X$}}(g_0)$ for all $g_0 \in \mathcal{G}_0$, but does not know the value $g_0$ chosen by the interfering user. Let vectors $\mbox{\boldmath  $g$}$ and $\mbox{\boldmath  $\mathcal{G}$}$ now contain the entry of the interfering user, while vectors $\mbox{\boldmath  $w$}$ and $\mbox{\boldmath  $X$}$ still only contain the entries of the regular users.

As in Section \ref{SectionII}, we assume that the receiver is only interested in decoding the message of user $1$. Let $(\mbox{\boldmath $w$},\mbox{\boldmath $g$})$ be the actual message vector and code index vector pair, unknown to the receiver. The receiver should choose an operation region $\mbox{\boldmath  $R$}_1$ in the space of $\mbox{\boldmath  $g$}$. The receiver intends to decode the message of user $1$ if $\mbox{\boldmath  $g$}\in \mbox{\boldmath  $R$}_1$, and intends to report collision for user $1$ if $\mbox{\boldmath  $g$}\not\in \mbox{\boldmath  $R$}_1$.

\begin{theorem}\label{MultipleAccessSingleDecodingInterferingCapacity}
	For a discrete-time memoryless multiple access channel $P_{Y|\mbox{\scriptsize \boldmath  $X$}}(g_0)$ with finite input and output alphabets and with $g_0$ being the code index of an interfering user, conclusions of Theorems \ref{MultipleAccessSingleDecodingAchievability}, \ref{MultipleAccessSingleDecodingCapacity}, and Corollaries \ref{MultipleAccessGroupDecodingCapacity} still hold, if the following extensions are applied to the statements in the theorems, corollaries and in their proofs.
	
	1. Channel input vectors $\mbox{\boldmath  $X$}$, rate vectors $\mbox{\boldmath  $r$}_{\mbox{\scriptsize \boldmath  $g$}}$, input distribution vectors $\mbox{\boldmath $P$}_{\mbox{\scriptsize \boldmath $gX$}}$ should only contain entries corresponding to the regular users $1, \dots, K$.
	
	2. Code index vectors $\mbox{\boldmath  $g$}=(\mbox{\boldmath  $r$}_{\mbox{\scriptsize \boldmath  $g$}}, \mbox{\boldmath $P$}_{\mbox{\scriptsize \boldmath $gX$}}, g_0)$ as well as code ensemble vector $\mbox{\boldmath  $\mathcal{G}$}$ should contain one more entry corresponding to the code index of the interfering user.
	
	3. Given code index vector $\mbox{\boldmath  $g$}$, mutual information function $I_{\mbox{\scriptsize \boldmath  $g$}}( )$, entropy function $H_{\mbox{\scriptsize \boldmath  $g$}}( )$, and probability function $p_{\mbox{\scriptsize \boldmath  $g$}}( )$ should all be computed with respect to joint distribution $P_{\mbox{\scriptsize \boldmath $X$}Y}=P_{Y|\mbox{\scriptsize \boldmath $X$}}(g_0)\prod_{k=1}^{K}P_{g_kX_k}$, i.e., with a channel function of $P_{Y|\mbox{\scriptsize \boldmath $X$}}(g_0)$.
	
	4. User subsets $S\subseteq\{1, \dots, K\}$ should only contain the regular users. The complement set $\bar{S}$ should be defined as $\bar{S}=\{1, \dots, K\} \setminus S$, i.e., excluding the interfering user.
	
	5. The maximum number of possible code index vectors should be upper bounded by $M^{K+1}$.
	
    With the above extensions, if error probability is defined in (\ref{MultipleAccessSingleDecodingErrorProbability}), then any subset of an achievable region should also be achievable. $C_{d1}$ given in (\ref{MultipleAccessUser1Capacity}) is the maximum asymptotically achievable region for user $1$, and $\mbox{\boldmath $C$}_{dS_0}$ given in (\ref{MultipleAccessUserS0Capacity}) is the maximum asymptotically achievable region for user subset $S_0 \subseteq \{1, \dots, K\}$.
\end{theorem}

The proof of Theorem \ref{MultipleAccessSingleDecodingInterferingCapacity} is skipped.

\section{Performance with A Finite Codeword Length}
\label{SectionIV}
Following the system model introduced in Section \ref{SectionIII}, in this section, we present the non-asymptotic analysis when the codeword length is finite and could be small in value. Throughout this section, codeword length $N$ is assumed to be fixed at a constant.

As explained in \cite{ref Luo15}, we will first need to consider an axillary decoder called the $(D, \mbox{\boldmath  $R$}_D)$ decoder. Let $D \subseteq \{1, \dots, K\}$ be a subset of regular users with $1\in D$. Assume that the receiver chooses an operation region $\mbox{\boldmath  $R$}_D$ and an operation margin $\mbox{\boldmath  $\widehat{R}$}_D$ both defined in the code space with $\mbox{\boldmath  $R$}_D \cap \mbox{\boldmath  $\widehat{R}$}_D=\phi$. A $(D, \mbox{\boldmath  $R$}_D)$ decoder intends to decode the messages of all users in $D$ by regarding signals from all other users as interference. Let $(\mbox{\boldmath  $w$}, \mbox{\boldmath  $g$})$ be the actual message vector and code index vector pair. For $\mbox{\boldmath  $g$}\in \mbox{\boldmath  $R$}_D$, the decoder intends to decode the messages of users in $D$. For $\mbox{\boldmath  $g$}\in \mbox{\boldmath  $\widehat{R}$}_D$, the decoder intends to either decode the messages or to report collision for users in $D$. For $\mbox{\boldmath  $g$}\not\in \mbox{\boldmath  $R$}_D\cup\mbox{\boldmath  $\widehat{R}$}_D$, the decoder intends to enforce collision report for users in $D$. Let $(\mbox{\boldmath  $\hat{w}$}_D, \mbox{\boldmath  $\hat{g}$}_D)$ be the estimated message vector and code index vector for users in $D$. Given $\mbox{\boldmath $g$}$, conditional error probability as a function of $\mbox{\boldmath $g$}$ is given by
\begin{eqnarray}
	&& P_e(\mbox{\boldmath $g$})= \Biggl \{ \nonumber \\
	&& \max_{\mbox{\scriptsize \boldmath $w$}_D}Pr\{(\mbox{\boldmath  $\hat{w}$}_D, \mbox{\boldmath  $\hat{g}$}_D)\ne (\mbox{\boldmath  $w$}_D, \mbox{\boldmath  $g$}_D)|(\mbox{\boldmath  $w$}_D, \mbox{\boldmath  $g$})\}, \forall \mbox{\boldmath $g$}\in \mbox{\boldmath $R$}_D \nonumber \\
 && \max_{\mbox{\scriptsize \boldmath $w$}_D} 1-Pr\left\{\left.\begin{array}{l}\mbox{``collision'' or}\\ (\mbox{\boldmath  $\hat{w}$}_D, \mbox{\boldmath  $\hat{g}$}_D)= (\mbox{\boldmath  $w$}_D, \mbox{\boldmath  $g$}_D) \end{array}\right|(\mbox{\boldmath  $w$}_D, \mbox{\boldmath  $g$}) \right\}, \nonumber \\
 && \qquad \qquad \qquad \qquad \qquad \qquad \qquad \qquad \quad \forall \mbox{\boldmath $g$}\in \widehat{\mbox{\boldmath $R$}}_D \nonumber \\
 && \max_{\mbox{\scriptsize \boldmath $w$}_D} 1-Pr\left\{\mbox{``collision''}|(\mbox{\boldmath  $w$}_D, \mbox{\boldmath  $g$}) \right\}, \forall \mbox{\boldmath $g$}\not\in \mbox{\boldmath $R$}_D \cup \widehat{\mbox{\boldmath $R$}}_D  \nonumber \\
	\label{DDecodingErrorProbability}
\end{eqnarray}

Let $\{\alpha_{\mbox{\scriptsize \boldmath $g$}}\}$ be a set of pre-determined weight parameters each being assigned to a code index vector $\mbox{\boldmath $g$} \in \mbox{\boldmath $\mathcal{G}$}$, such that
\begin{equation}
	\left\{\alpha_{\mbox{\scriptsize \boldmath $g$}}\left| \alpha_{\mbox{\scriptsize \boldmath $g$}} \ge 0, \forall \mbox{\boldmath $g$} \in \mbox{\boldmath $\mathcal{G}$}, \sum_{\mbox{\scriptsize \boldmath $g$}} e^{-N\alpha_{\mbox{\scriptsize \boldmath $g$}}}=1 \right.\right\}.
	\label{WeightParamater}
\end{equation}
We define the ``generalized error performance'' of the $(D, \mbox{\boldmath  $R$}_D)$ decoder as
\begin{equation}
\mbox{GEP}_D=\sum_{\mbox{\scriptsize \boldmath $g$}}P_e(\mbox{\boldmath $g$})e^{-N\alpha_{\mbox{\scriptsize \boldmath $g$}}}.
\end{equation}

Let us use $P_{g_k}(X_k)$ to denote the probability of channel input symbol $X_k$ under coding option $g_k$, and use $P(Y|\mbox{\boldmath $X$}_D, \mbox{\boldmath $g$}_{\bar{D}})$ to denote the conditional probability of channel output symbol $Y$ given input symbol vector $\mbox{\boldmath $X$}_D$ for users in $D$, and code index vector $\mbox{\boldmath $g$}_{\bar{D}}$ for users not in $D$. The following theorem gives an achievable bound, improved from the corresponding bound presented in \cite[Theorem 3]{ref Luo15}, for the generalized error performance of the $(D, \mbox{\boldmath  $R$}_D)$ decoder.

\begin{theorem}\label{MultipleAccessGeneralErrorPerformanceBound}
	Consider the distributed multiple access system described above. There exists a decoding algorithm such that $\mbox{GEP}_D$ is upper bounded by
	\begin{eqnarray}
		&& \mbox{GEP}_D \le\nonumber \\
		&&\sum_{\mbox{\scriptsize \boldmath $g$}\in \mbox{\scriptsize \boldmath $R$}_D}\left\{\sum_{S \subset D} \left[ \sum_{\tilde{\mbox{\scriptsize \boldmath $g$}}\in \mbox{\scriptsize \boldmath $R$}_D, \tilde{\mbox{\scriptsize \boldmath $g$}}_S= \mbox{\scriptsize \boldmath $g$}_S} \exp(-N E_{mD}(\mbox{\boldmath $g$},\tilde{\mbox{\boldmath $g$}}, S)) \right. \right. \nonumber \\
		&& \quad + 2 \sum_{\tilde{\mbox{\scriptsize \boldmath $g$}}\not\in \mbox{\scriptsize \boldmath $R$}_D, \tilde{\mbox{\scriptsize \boldmath $g$}}_S= \mbox{\scriptsize \boldmath $g$}_S}  \exp(-N E_{iD}(\mbox{\boldmath $g$}, \tilde{\mbox{\boldmath $g$}}, S))\Biggr] \nonumber \\
		&& \quad \left. + 2 \sum_{\tilde{\mbox{\scriptsize \boldmath $g$}}\not\in \mbox{\scriptsize \boldmath $R$}_D \cup \widehat{\mbox{\scriptsize \boldmath $R$}}_D, \tilde{\mbox{\scriptsize \boldmath $g$}}_D= \mbox{\scriptsize \boldmath $g$}_D}  \exp(-N E_{iD}(\mbox{\boldmath $g$}, \tilde{\mbox{\boldmath $g$}}, D))\right\},
		\label{MultipleAccessGeneralGEPAchievabilityBound}
	\end{eqnarray}
	where $E_{mD}(\mbox{\boldmath $g$},\tilde{\mbox{\boldmath $g$}}, S)$, $E_{iD}(\mbox{\boldmath $g$}, \tilde{\mbox{\boldmath $g$}}, S)$ for $S\subset D$ and $E_{iD}(\mbox{\boldmath $g$}, \tilde{\mbox{\boldmath $g$}}, D)$ in the above equation are given by

	\begin{eqnarray}
		&& E_{mD}(\mbox{\boldmath $g$},\tilde{\mbox{\boldmath $g$}}, S)= \nonumber \\
		&& \max_{0<\rho\le 1}-\rho \sum_{k\in D\setminus S} r_{\tilde{g}_k} +\max_{0\le s\le 1}-\log \sum_Y\sum_{\mbox{\scriptsize \boldmath $X$}_S} \prod_{k\in S}P_{g_k}(X_k)  \nonumber \\
		&& \times \left( \sum_{\mbox{\scriptsize \boldmath $X$}_{D\setminus S}}\left[P(Y|\mbox{\boldmath $X$}_D, \mbox{\boldmath $g$}_{\bar{D}})e^{-\alpha_{\mbox{\scriptsize \boldmath $g$}}}\right]^{1-s} \prod_{k\in D\setminus S}P_{g_k}(X_k) \right) \nonumber \\
		&& \times \left(\sum_{\mbox{\scriptsize \boldmath $X$}_{D\setminus S}} \left[P(Y|\mbox{\boldmath $X$}_D, \tilde{\mbox{\boldmath $g$}}_{\bar{D}})e^{-\alpha_{\tilde{\mbox{\scriptsize \boldmath $g$}}}}\right]^{\frac{s}{\rho}} \prod_{k\in D\setminus S}P_{\tilde{g}_k}(X_k)\right)^{\rho}, \nonumber \\
		&& E_{iD}(\mbox{\boldmath $g$}, \tilde{\mbox{\boldmath $g$}}, S)= \max_{0<\rho\le 1}-\rho \sum_{k\in D\setminus S} r_{g_k} \nonumber \\
		&& +\max_{0\le s\le 1-\rho}-\log \sum_Y\sum_{\mbox{\scriptsize \boldmath $X$}_S}\prod_{k\in S}P_{g_k}(X_k) \times   \nonumber \\
		&& \left( \sum_{\mbox{\scriptsize \boldmath $X$}_{D\setminus S}}\left[P(Y|\mbox{\boldmath $X$}_D, \mbox{\boldmath $g$}_{\bar{D}})e^{-\alpha_{\mbox{\scriptsize \boldmath $g$}}}\right]^{\frac{s}{s+\rho}} \prod_{k\in D\setminus S}P_{g_k}(X_k) \right)^{s+\rho} \nonumber \\
		&& \times \left(  \sum_{\mbox{\scriptsize \boldmath $X$}_{D\setminus S}} P(Y|\mbox{\boldmath $X$}_D, \tilde{\mbox{\boldmath $g$}}_{\bar{D}})e^{-\alpha_{\tilde{\mbox{\scriptsize \boldmath $g$}}}}  \prod_{k\in D\setminus S}P_{\tilde{g}_k}(X_k)  \right)^{1-s}. \nonumber \\
		&& E_{iD}(\mbox{\boldmath $g$}, \tilde{\mbox{\boldmath $g$}}, D)= \max_{0\le s\le 1}-\log \sum_Y\sum_{\mbox{\scriptsize \boldmath $X$}_D}\prod_{k\in D}P_{g_k}(X_k) \nonumber \\
		&& \left[P(Y|\mbox{\boldmath $X$}_D, \mbox{\boldmath $g$}_{\bar{D}})e^{-\alpha_{\mbox{\scriptsize \boldmath $g$}}}\right]^{s} \left[   P(Y|\mbox{\boldmath $X$}_D, \tilde{\mbox{\boldmath $g$}}_{\bar{D}})e^{-\alpha_{\tilde{\mbox{\scriptsize \boldmath $g$}}}}  \right]^{1-s}.
		\label{MultipleAccessGaneralErrorExponent}
	\end{eqnarray}
\end{theorem}

The proof of Theorem \ref{MultipleAccessGeneralErrorPerformanceBound} is given in Appendix \ref{ProofofMultipleAccessGeneralErrorPerformanceBound}. Compared with the bound presented in \cite[Equation (7)]{ref Luo15}, besides other minor improvements, the second and the third terms on the right hand side of (\ref{MultipleAccessGeneralGEPAchievabilityBound}) lead to a tighter bound because, if the summations are dominated by only a small number of terms, then the summations should not scale in the number of code index vectors satisfying $\tilde{\mbox{\boldmath $g$}}\not\in \mbox{\boldmath $R$}_D$.

Let us now consider the case when the receiver is only interested in decoding the message of user $1$ but can choose to decode the messages of other users if necessary. Assume that the receiver should choose an operation region $\mbox{\boldmath $R$}_1$ and an operation margin $\mbox{\boldmath $\widehat{R}$}_1$ in the code space with $\mbox{\boldmath $R$}_1 \cap \mbox{\boldmath $\widehat{R}$}_1=\phi$. Let $\mbox{\boldmath  $g$}$ be the actual code index vector. The receiver intends to decode the message of user $1$ for $\mbox{\boldmath  $g$}\in \mbox{\boldmath $R$}_1$, to either decode the message of user $1$ or to report collision for user $1$ for $\mbox{\boldmath  $g$}\in\mbox{\boldmath $\widehat{R}$}_1$, and to report collision for user $1$ for $\mbox{\boldmath  $g$}\not\in \mbox{\boldmath $R$}_1 \cup \mbox{\boldmath $\widehat{R}$}_1$.

Let $(\hat{w}_1, \hat{g}_1)$ be the message and code index estimate of user $1$. Let $(\mbox{\boldmath  $w$}, \mbox{\boldmath  $g$})$ be the actual message vector and code index vector pair, conditional error probability of the system as a function of $\mbox{\boldmath  $g$}$ is defined as
\begin{eqnarray}
	&& P_e(\mbox{\boldmath $g$})= \Biggl \{ \nonumber \\
	&& \max_{w_1}Pr\{(\hat{w}_1, \hat{g}_1)\ne (w_1, g_1)|(w_1, \mbox{\boldmath  $g$})\}, \forall \mbox{\boldmath $g$}\in \mbox{\boldmath $R$}_1 \nonumber \\
 && \max_{w_1} 1-Pr\left\{\left.\begin{array}{l}\mbox{``collision'' or}\\ (\hat{w}_1, \hat{g}_1)= (w_1, g_1)|(w_1, \mbox{\boldmath  $g$}) \end{array}\right|(w_1, \mbox{\boldmath  $g$}) \right\}, \nonumber \\
 && \qquad \qquad \qquad \qquad \qquad \qquad \qquad \qquad \quad \forall \mbox{\boldmath $g$}\in \widehat{\mbox{\boldmath $R$}}_1 \nonumber \\
 && \max_{w_1} 1-Pr\left\{\mbox{``collision''}|(w_1, \mbox{\boldmath  $g$}) \right\}, \forall \mbox{\boldmath $g$}\not\in \mbox{\boldmath $R$}_1 \cup \widehat{\mbox{\boldmath $R$}}_1.
 	\label{SingleDecodingErrorProbability}
\end{eqnarray}
Let $\{\alpha_{\mbox{\scriptsize \boldmath $g$}}\}$ be a set of pre-determined weight parameters each being assigned to a code index vector $\mbox{\boldmath $g$} \in \mbox{\boldmath $\mathcal{G}$}$ and satisfying constraint (\ref{WeightParamater}). We define the ``generalized error performance'' of the system as
\begin{equation}
	\mbox{GEP}=\sum_{\mbox{\scriptsize \boldmath $g$}}P_e(\mbox{\boldmath $g$})e^{-N\alpha_{\mbox{\scriptsize \boldmath $g$}}}.
	\label{SingleGeneralizedError}
\end{equation}

According to \cite[Theorem 4]{ref Luo15}, an achievable bound on the generalized error performance of the system is given in the following theorem.

\begin{theorem}\label{MultipleAccessSingleDecodingErrorPerformanceBound}
	Consider the distributed multiple access system described above. Assume that the receiver is only interested in decoding the message of user $1$. Let $\mbox{\boldmath $R$}_1$ be the operation region, $\widehat{\mbox{\boldmath $R$}}_1$ be the operation margin, and $\{\alpha_{\mbox{\scriptsize \boldmath $g$}}\}$ be the set of weight parameters. Let $\sigma$ be a partition of the operation region $\mbox{\boldmath $R$}_1$, as described below
	\begin{eqnarray}
		&& \mbox{\boldmath $R$}_1=\bigcup_{D, D\subseteq \{1, \dots, K\}, 1\in D} \mbox{\boldmath $R$}_D, \qquad \quad \mbox{\boldmath $R$}_{D'}\cap \mbox{\boldmath $R$}_{D}=\phi, \nonumber \\
		&& \forall D, D' \subseteq \{1, \dots, K\}, D'\ne D, 1\in D, D'.
	\end{eqnarray}
	There exists a decoding algorithm such that the generalized error performance defined in (\ref{SingleGeneralizedError}) is upper bounded by
	\begin{equation}
		\mbox{GEP}\le \min_{\sigma} \sum_{D,D\subseteq \{1, \dots, K\}, 1\in D} \mbox{GEP}_D,
		\label{MultipleAccessSingleDecodingGeneralGEPAchievabilityBound}
	\end{equation}
	where $\mbox{GEP}_D$ represents the generalized error probability of the $(D, \mbox{\boldmath $R$}_D)$ decoder with receiver decoding the messages of all and only the users in $D$, with the operation region being $\mbox{\boldmath $R$}_D$ and the operation margin being $\widehat{\mbox{\boldmath $R$}}_D=\mbox{\boldmath $R$}_1\cup \widehat{\mbox{\boldmath $R$}}_1 \setminus \mbox{\boldmath $R$}_D$.
\end{theorem}

\appendix

\subsection{Proof of Theorem \ref{MultipleAccessSingleDecodingCapacity}}
\label{ProofofMultipleAccessSingleDecodingCapacity}

\begin{proof} Achievability part of the theorem is implied by \cite[Theorem 1]{ref Luo15}. To prove the converse part, consider an operation region $\mbox{\boldmath $R$}_1$ that is asymptotically achievable for user $1$. Let $\mbox{\boldmath $g$}\in\mbox{\boldmath $R$}_1$ be an arbitrary code index vector in $\mbox{\boldmath $R$}_1$. We will show that $\mbox{\boldmath $g$}\in\mbox{\boldmath $C$}_{d1}^c$ must be true.

Let $(\mbox{\boldmath $w$}, \mbox{\boldmath $g$})$ be the actual message vector and code index vector pair. We assume $\mbox{\boldmath $g$}$ is known to the receiver. We will also skip $\mbox{\boldmath $g$}$ in the subscription to simplify the notation. Since $\mbox{\boldmath $g$}\in \mbox{\boldmath $R$}_1$, the receiver should output $\hat{w}_1=w_1$ with an asymptotic probability of one. Let $S\subseteq \{1, \dots, K\}$ be an arbitrary user subset with $1\in S$. Assume that codewords of users in $\bar{S}$ are known at the receiver. Because the message of user $1$ is correctly decoded with an asymptotic probability of one, there must exist a user subset $\tilde{S} \subseteq S$ with $1\in \tilde{S}$ such that, with an asymptotic probability of one, the receiver can jointly decode the messages of users in $\tilde{S}$ by regarding the input symbols from users in $S\setminus \tilde{S}$ as interference. Denote the probability that the receiver is not able to recover the messages of all users in $\tilde{S}$ as $P_e^{(N)}(\tilde{S})$, we have $\lim_{N\to\infty}P_e^{(N)}(\tilde{S})=0$.

Let $\epsilon>0$ be an arbitrary small constant. According to Fano's inequality, for large enough $N$, we have
\begin{eqnarray}
&& \sum_{k\in \tilde{S}} r_{g_k} \le \frac{1}{N}H(\mbox{\boldmath $w$}_{\tilde{S}})+\epsilon =\frac{1}{N}H\left(\mbox{\boldmath $w$}_{\tilde{S}}|\mbox{\boldmath $X$}_{\bar{S}}^{(N)}(\mbox{\boldmath $w$}_{\bar{S}})\right) +\epsilon \nonumber \\
&& = \frac{1}{N} H\left(\mbox{\boldmath $w$}_{\tilde{S}}|\mbox{\boldmath $X$}_{\bar{S}}^{(N)}(\mbox{\boldmath $w$}_{\bar{S}}), Y^{(N)}\right) \nonumber \\
&& \quad +\frac{1}{N}I\left(\mbox{\boldmath $w$}_{\tilde{S}}; Y^{(N)}|\mbox{\boldmath $X$}_{\bar{S}}^{(N)}(\mbox{\boldmath $w$}_{\bar{S}}) \right) +\epsilon \nonumber \\
&& \le \frac{1}{N}I\left(\mbox{\boldmath $w$}_{\tilde{S}}; Y^{(N)}|\mbox{\boldmath $X$}_{\bar{S}}^{(N)}(\mbox{\boldmath $w$}_{\bar{S}}) \right) +2\epsilon \nonumber \\
&& \le \frac{1}{N}I\left(\mbox{\boldmath $X$}_{\tilde{S}}^{(N)}; Y^{(N)}|\mbox{\boldmath $X$}_{\bar{S}}^{(N)}(\mbox{\boldmath $w$}_{\bar{S}}) \right) +2\epsilon \nonumber \\
&& = I\left(\mbox{\boldmath $X$}_{\tilde{S}}; Y|\mbox{\boldmath $X$}_{\bar{S}} \right) +2\epsilon,
\label{MultipleAccessSingleDecodingFanoBound}
\end{eqnarray}
where the last equality is due to the fact that the channel is memoryless and codeword symbols are generated independently. By taking $N$ to infinity and taking $\epsilon$ to $0$, (\ref{MultipleAccessSingleDecodingFanoBound}) implies that $\sum_{k\in \tilde{S}} r_{g_k} \le I\left(\mbox{\boldmath $X$}_{\tilde{S}}; Y|\mbox{\boldmath $X$}_{\bar{S}} \right)$. Since this holds for every user subset $S\subseteq \{1, \dots, K\}$ with $1\in S$, we must have $\mbox{\boldmath $g$}\in \mbox{\boldmath $C$}_{d1}^c$. Because $\mbox{\boldmath $g$}\in \mbox{\boldmath $R$}_1$ is chosen arbitrarily, $\mbox{\boldmath $R$}_1 \subseteq \mbox{\boldmath $C$}_{d1}^c$ therefore must be true.
\end{proof}

\subsection{Proof of Theorem \ref{MultipleAccessGeneralErrorPerformanceBound}}
\label{ProofofMultipleAccessGeneralErrorPerformanceBound}

\begin{proof}
Given channel output sequence $Y^{(N)}$, channel input vector sequence $\mbox{\boldmath $X$}_D^{(N)}$ and code index vector $\mbox{\boldmath $g$}$, we define the weighted likelihood of the channel input sequence as $L_{\mbox{\scriptsize \boldmath $g$}}\left(\mbox{\boldmath $X$}_D^{(N)},Y^{(N)}\right)=P(Y^{(N)}|\mbox{\boldmath $X$}_D^{(N)}, \mbox{\boldmath $g$}_{\bar{D}})e^{-N\alpha_{\mbox{\scriptsize \boldmath $g$}}}$.

For every user subset $S \subseteq D$, we define a constraint set $\mbox{\boldmath $\mathcal{R}$}_S$ of message vector and code index vector pairs. Each code index vector in the constraint set should belong to the operation region and weighted likelihood of the corresponding codeword vector should stay above a pre-determined threshold.
\begin{eqnarray}
&& \mbox{\boldmath $\mathcal{R}$}_S=\biggl\{(\mbox{\boldmath $w$}_D, \mbox{\boldmath $g$}) \left|  \mbox{\boldmath $g$}\in \mbox{\boldmath $R$}_D, \forall  \tilde{\mbox{\boldmath $g$}}\not\in \mbox{\boldmath $R$}_D \mbox{ with } \mbox{\boldmath $g$}_S=\tilde{\mbox{\boldmath $g$}}_S, \right. \nonumber \\
&& \qquad \left. L_{\mbox{\scriptsize \boldmath $g$}}\left(\mbox{\boldmath $X$}_D^{(N)},Y^{(N)}\right)> e^{-N\tau_{[\mbox{\scriptsize \boldmath $g$}, \tilde{\mbox{\scriptsize \boldmath $g$}}, S]}(\mbox{\scriptsize \boldmath $X$}_S^{(N)}, Y^{(N)})} \right\},\nonumber \\
\label{TaugSXSY}
\end{eqnarray}
where $\tau_{[\mbox{\scriptsize \boldmath $g$}, \tilde{\mbox{\scriptsize \boldmath $g$}}, S]}(\mbox{\boldmath $X$}_S^{(N)}, Y^{(N)})$ is a threshold function whose value will be determined later. We further define constraint set $\mbox{\boldmath $\mathcal{R}$}_{P}=\bigcap_{S \subseteq D} \mbox{\boldmath $\mathcal{R}$}_S$ as the intersection of $\mbox{\boldmath $\mathcal{R}$}_{S}$ for all $S\subseteq D$.

Assume the following decoding algorithm at the receiver. Given $Y^{(N)}$, the receiver first calculates constraint sets $\mbox{\boldmath $\mathcal{R}$}_S$ for all $S \subseteq D$ to obtain constraint set $\mbox{\boldmath $\mathcal{R}$}_{P}$. The receiver reports collision for all users in $D$ if $\mbox{\boldmath $\mathcal{R}$}_{P}$ is empty. Otherwise, the receiver outputs $(\hat{\mbox{\boldmath $w$}}_D, \hat{\mbox{\boldmath $g$}})\in \mbox{\boldmath $\mathcal{R}$}_{P}$ with the maximum weighted likelihood value.

We define the notation $(\mbox{\boldmath $w$}_D, \mbox{\boldmath $g$}) \stackrel{S}{=} (\tilde{\mbox{\boldmath $w$}}_D, \tilde{\mbox{\boldmath $g$}})$ as
\begin{eqnarray}
&& (\mbox{\boldmath $w$}_D,\mbox{\boldmath $g$}) \stackrel{S}{=} (\tilde{\mbox{\boldmath $w$}}_D,\tilde{\mbox{\boldmath$g$}}) :      (\mbox{\boldmath $w$}_S,\mbox{\boldmath $g$}_S) = (\tilde{\mbox{\boldmath $w$}}_S, \tilde{\mbox{\boldmath $g$}}_S), \nonumber \\
&& \qquad \qquad (w_k,g_k)\neq(\tilde{w}_{k},\tilde{g}_k), \forall k \in D \setminus S.
\end{eqnarray}
$(\mbox{\boldmath $w$}_D, \mbox{\boldmath $g$}) \stackrel{S}{=} (\tilde{\mbox{\boldmath $w$}}_D, \tilde{\mbox{\boldmath $g$}})$ means that the two message vector and code index vector pairs are equal for users in $S$ and are different for users in $D\setminus S$. The term does not imply any assumption on code indices of the other users.

Assume that $(\mbox{\boldmath $w$}_D, \mbox{\boldmath $g$})$ with $\mbox{\boldmath $g$}\in \mbox{\boldmath $R$}_D$ is the actual message vector and code index vector pair. For any user subset $S \subset D$, we define $P_{m[\mbox{\scriptsize \boldmath $g$},\tilde{\mbox{\scriptsize \boldmath $g$}}, S]}$ as
\begin{eqnarray}
&& P_{m[\mbox{\scriptsize \boldmath $g$},\tilde{\mbox{\scriptsize \boldmath $g$}}, S]}=Pr\Bigl\{\exists \tilde{\mbox{\boldmath $w$}}_D, (\tilde{\mbox{\boldmath $w$}}_D, \tilde{\mbox{\boldmath $g$}}) \stackrel{S}{=} (\mbox{\boldmath $w$}_D, \mbox{\boldmath $g$}), \mbox{ such that } \nonumber \\
&& \left. L_{\mbox{\scriptsize \boldmath $g$}}\left(\mbox{\boldmath $X$}_{\mbox{\scriptsize \boldmath $g$}_D}^{(N)}(\mbox{\boldmath $w$}_D),Y^{(N)}\right) \le L_{\tilde{\mbox{\scriptsize \boldmath $g$}}}\left(\mbox{\boldmath $X$}_{\tilde{\mbox{\scriptsize \boldmath $g$}}_D}^{(N)}(\tilde{\mbox{\boldmath $w$}}_D),Y^{(N)}\right) \right\} \nonumber \\
&& \qquad \qquad \qquad  \mbox{ for } \mbox{\boldmath $g$}, \tilde{\mbox{\boldmath $g$}}\in \mbox{\boldmath $R$}_D \mbox{ with } \mbox{\boldmath $g$}_S=\tilde{\mbox{\boldmath $g$}}_S.
\label{PmgtildegSDefinition}
\end{eqnarray}
For any user subset $S\subseteq D$, we define $P_{t[\mbox{\scriptsize \boldmath $g$}, \tilde{\mbox{\scriptsize \boldmath $g$}}, S]}$ as
\begin{eqnarray}
&& P_{t[\mbox{\scriptsize \boldmath $g$}, \tilde{\mbox{\scriptsize \boldmath $g$}}, S]}=Pr\Bigl\{ \nonumber \\
&& \left. L_{\mbox{\scriptsize \boldmath $g$}}\left(\mbox{\boldmath $X$}_{\mbox{\scriptsize \boldmath $g$}_D}^{(N)}(\mbox{\boldmath $w$}_D),Y^{(N)}\right) \le e^{-N\tau_{[\mbox{\scriptsize \boldmath $g$}, \tilde{\mbox{\scriptsize \boldmath $g$}}, S]}(\mbox{\scriptsize \boldmath $X$}_S^{(N)}, Y^{(N)})} \right\} \nonumber \\
&& \qquad \qquad \mbox{ for } \mbox{\boldmath $g$}\in \mbox{\boldmath $R$}_D, \tilde{\mbox{\boldmath $g$}}\not\in \mbox{\boldmath $R$}_D \mbox{ with } \mbox{\boldmath $g$}_S=\tilde{\mbox{\boldmath $g$}}_S.
\label{PtgSDefinition}
\end{eqnarray}

Assume that $(\tilde{\mbox{\boldmath $w$}}_D, \tilde{\mbox{\boldmath $g$}})$ with $\tilde{\mbox{\boldmath $g$}}\not\in \mbox{\boldmath $R$}_D$ is the actual message vector and code index vector pair. For any user subset $S \subseteq D$, we define $P_{i[\tilde{\mbox{\scriptsize \boldmath $g$}}, \mbox{\scriptsize \boldmath $g$}, S]}$ as
\begin{eqnarray}
&& P_{i[\tilde{\mbox{\scriptsize \boldmath $g$}}, \mbox{\scriptsize \boldmath $g$}, S]}=Pr\left\{\exists \tilde{\mbox{\boldmath $w$}}_D, (\tilde{\mbox{\boldmath $w$}}_D, \tilde{\mbox{\boldmath $g$}}) \stackrel{S}{=} (\mbox{\boldmath $w$}_D, \mbox{\boldmath $g$}), \mbox{ such that } \right.\nonumber \\
&& \left. L_{\mbox{\scriptsize \boldmath $g$}}\left(\mbox{\boldmath $X$}_{\mbox{\scriptsize \boldmath $g$}_D}^{(N)}(\mbox{\boldmath $w$}_D),Y^{(N)}\right) > e^{-N\tau_{[\mbox{\scriptsize \boldmath $g$}, \tilde{\mbox{\scriptsize \boldmath $g$}}, S]}(\mbox{\scriptsize \boldmath $X$}_S^{(N)}, Y^{(N)})} \right\} \nonumber \\
&& \qquad \qquad \mbox{ for } \tilde{\mbox{\boldmath $g$}} \not\in \mbox{\boldmath $R$}_D, \mbox{\boldmath $g$}\in \mbox{\boldmath $R$}_D \mbox{ with } \mbox{\boldmath $g$}_S=\tilde{\mbox{\boldmath $g$}}_S.
\label{PitildeggSDefinition}
\end{eqnarray}

With the above probability definitions, the generalized error performance of the system can be upper bounded by
\begin{eqnarray}
&& \mbox{GEP}_D\le  \sum_{\mbox{\scriptsize \boldmath $g$}\in \mbox{\scriptsize \boldmath $R$}_D} \left\{\sum_{S \subset D}\left[ \sum_{ \tilde{\mbox{\scriptsize \boldmath $g$}}\in \mbox{\scriptsize \boldmath $R$}_D, \tilde{\mbox{\scriptsize \boldmath $g$}}_S= \mbox{\scriptsize \boldmath $g$}_S}P_{m[\mbox{\scriptsize \boldmath $g$},\tilde{\mbox{\scriptsize \boldmath $g$}}, S]}e^{-N\alpha_{\mbox{\scriptsize \boldmath $g$}}}  \right. \right. \nonumber \\
&& \left. +\sum_{\tilde{\mbox{\scriptsize \boldmath $g$}}\not\in \mbox{\scriptsize \boldmath $R$}_D, \tilde{\mbox{\scriptsize \boldmath $g$}}_S= \mbox{\scriptsize \boldmath $g$}_S}\left(P_{t[\mbox{\scriptsize \boldmath $g$},\tilde{\mbox{\scriptsize \boldmath $g$}}, S]}e^{-N\alpha_{\mbox{\scriptsize \boldmath $g$}}}+P_{i[\tilde{\mbox{\scriptsize \boldmath $g$}}, \mbox{\scriptsize \boldmath $g$}, S]}e^{-N\alpha_{\tilde{\mbox{\scriptsize \boldmath $g$}}}}\right)\right]+ \nonumber \\
&& \left. \sum_{ \stackrel{\tilde{\mbox{\scriptsize \boldmath $g$}}\not\in \mbox{\scriptsize \boldmath $R$}_D\cup \widehat{\mbox{\scriptsize \boldmath $R$}}_D ,}{\tilde{\mbox{\scriptsize \boldmath $g$}}_D= \mbox{\scriptsize \boldmath $g$}_D}}\left(P_{t[\mbox{\scriptsize \boldmath $g$},\tilde{\mbox{\scriptsize \boldmath $g$}}, D]}e^{-N\alpha_{\mbox{\scriptsize \boldmath $g$}}} + P_{i[\tilde{\mbox{\scriptsize \boldmath $g$}}, \mbox{\scriptsize \boldmath $g$}, D]}e^{-N\alpha_{\tilde{\mbox{\scriptsize \boldmath $g$}}}} \right)\right\}.
\label{GEPMultipleAccessUpperBound}
\end{eqnarray}

By following Step I in the proof of \cite[Theorem 3]{ref Luo15}, we get
\begin{equation}
P_{m[\mbox{\scriptsize \boldmath $g$},\tilde{\mbox{\scriptsize \boldmath $g$}}, S]}e^{-N\alpha_{\mbox{\scriptsize \boldmath $g$}}} \le \exp(-N E_{mD}(\mbox{\boldmath $g$},\tilde{\mbox{\boldmath $g$}}, S)).
\label{MultipleAccessPmgFinalBound}
\end{equation}
By following Steps II, III, IV in the proof of \cite[Theorem 3]{ref Luo15} with minor modifications, we get
\begin{eqnarray}
&& P_{t[\mbox{\scriptsize \boldmath $g$}, \tilde{\mbox{\scriptsize \boldmath $g$}}, S]}e^{-N\alpha_{\mbox{\scriptsize \boldmath $g$}}} \le \exp(-NE_{iD}(\mbox{\boldmath $g$}, \tilde{\mbox{\boldmath $g$}}, S)) \nonumber \\
&& P_{i[\tilde{\mbox{\scriptsize \boldmath $g$}}, \mbox{\scriptsize \boldmath $g$}, S]}e^{-N\alpha_{\tilde{\mbox{\scriptsize \boldmath $g$}}}} \le \exp(-NE_{iD}(\mbox{\boldmath $g$}, \tilde{\mbox{\boldmath $g$}}, S)).
\label{MultipleAccessPtgigFinalBound}
\end{eqnarray}
Conclusion of the theorem then follows.

Note that, compared with the proof of \cite[Theorem 3]{ref Luo15}, the key revision here is the introduction of $\tau_{[\mbox{\scriptsize \boldmath $g$}, \tilde{\mbox{\scriptsize \boldmath $g$}}, S]}(\mbox{\boldmath $X$}_S^{(N)}, Y^{(N)})$ for each pair of code index vectors $\mbox{\boldmath $g$}$ and $\tilde{\mbox{\boldmath $g$}}$. This is opposed to using only one threshold variable for each $\mbox{\boldmath $g$} \in \mbox{\boldmath $R$}_D$, as suggested in \cite[Theorem 3]{ref Luo15}.

\end{proof}



\begin{thebibliography}{1}
	
\bibitem{ref Luo12}
J. Luo and A. Ephremides, ``A New Approach to Random Access: Reliable Communication and Reliable Collision Detection,''
\emph{IEEE Trans. on Inform. Theory}, Vol. 58, pp. 379-423, Feb. 2012.
	
\bibitem{ref Wang12}
Z. Wang and J. Luo, ``Error Performance of Channel Coding in Random Access Communication,''
\emph{IEEE Trans. on Inform. Theory}, Vol. 58, pp. 3961-3974, Jun. 2012.
	
\bibitem{ref Luo15}
J. Luo, ``A Generalized Channel Coding Theory for Distributed Communication,''
\emph{IEEE Trans. on Commun.}, Vol. 63, pp. 1043-1056, Apr. 2015.

\end{thebibliography}

\end{document}